# Custom silicon technology for SPAD-arrays with red-enhanced sensitivity and low timing jitter

**Angelo Gulinatti,[1,*] Francesco Ceccarelli,[1,2] Massimo Ghioni,[1] and Ivan Rech[1]**

[1]*Dipartimento di Elettronica, Informazione e Bioingegneria, Politecnico di Milano, piazza Leonardo da Vinci 32, 20133 Milano, Italy*
[2]*Currently with Istituto di Fotonica e Nanotecnologie – Consiglio Nazionale delle Ricerche (IFN-CNR) and Dipartimento di Fisica – Politecnico di Milano, piazza Leonardo da Vinci 32, 20133 Milano, Italy*
*\*angelo.gulinatti@polimi.it*

**Abstract:** Single-photon detection is an invaluable tool for many applications ranging from basic research to consumer electronics. In this respect, the Single Photon Avalanche Diode (SPAD) plays a key role in enabling a broad diffusion of these techniques thanks to its remarkable performance, room-temperature operation, and scalability. In this paper we present a silicon technology that allows the fabrication of SPAD-arrays with an unprecedented combination of low timing jitter (95 ps FWHM) and high detection efficiency at red and near infrared wavelengths (peak of 70% at 650 nm, 45% at 800 nm). We discuss the device structure, the fabrication process, and we present a thorough experimental characterization of the fabricated detectors. We think that these long-awaited results can pave the way to new exciting developments in many fields, ranging from quantum optics to single molecule spectroscopy.

## 1. Introduction

The capability of detecting single photons is a key requirement in an ever-increasing number of applications that span from fundamental research to consumer electronics. For example, single-photon detection is routinely employed in: *biology and biochemistry*, to detect the molecular environment, the interactions, and the conformational changes either at single-molecule level [1, 2] or across extended samples [3]; in *medical imaging*, to measure tissues properties through non-invasive techniques like Diffuse Optical Tomography [4] or Diffuse Correlation Spectroscopy [5]; in *autonomous driving*, for three-dimensional reconstruction of obstacles through Light Detection and Ranging (LiDAR) [6]; in *communications,* to secure data exchange by exploiting Quantum Key Distribution (QKD) [7].

Today, two different detectors technologies dominate the field of single photon counting: the Superconducting Nanowire Single Photon Detectors (SNSPDs), and the Single Photon Avalanche Diodes (SPADs). The formers are almost-ideal single-photon detectors, with detection efficiency exceeding 90% [8], spectral sensitivity that can extend in the mid-infrared [9], negligible dark count rate [10], and a timing jitter lower than 10 ps [11]. However, it is worth noting that these record figures-of-merit cannot be achieved simultaneously, because they are subjected to trades-off in detector design and operating conditions. Even more important, SNSPDs must be operated at cryogenic temperatures, typically between 1 and 4 K. Despite the incredible progresses in closed-cycle cryostats [12, 13], the need for cryo-cooling limits the use of these detectors to specific applications in which the benefits coming from their exceptional performance justify the cost, size, and complexity of such a system.

By contrast, SPADs are semiconductor devices usually operated at room temperature or moderately cooled (e.g. at -10 °C). This, in combination with their good performance, ruggedness, and scalability, makes the SPAD the detector of choice for a wide diffusion of the aforementioned single-photon applications. In these devices, single photons are detected by exploiting self-sustained avalanche multiplication in semiconductors. In particular, when a pn junction is biased above the breakdown voltage, initially, no current flows due to the absence of free carriers. In these conditions, even the electron-hole pair generated by the absorption of a single photon can trigger a self-sustained avalanche process, and the corresponding macroscopic current can be easily detected by a readout electronics. However, at this point the detector is blind until a suitable circuit turns the avalanche off and restores the initial biasing conditions [14]. This can happen in a time as short as a few nanoseconds [15].

In the last years, many researchers focused on CMOS SPADs, i.e. on SPADs fabricated by using the very same process-flow adopted for the manufacturing of Complementary Metal Oxide Semiconductor (CMOS) integrated circuits. Among the numerous advantages of this approach, there is the possibility of building both detectors and high-performance electronics on the same silicon chip. This led to the development of smart pixels with advanced functionalities [16] and to the fabrication of large arrays of SPADs [17, 18]. A drawback of the CMOS approach is in the limited degrees of freedom that it offers for the design of the detector; this constraint stems from the impossibility of modifying a process flow usually optimized for transistor performance.

A field in which this limitation is especially important is the development of SPADs with high detection efficiency and low timing jitter in the red and near infrared region of the spectrum. These are key requirements for a broad class of applications, ranging from single-molecules spectroscopy [19] to satellite-based Quantum Key Distribution (QKD) [20, 21]. Actually, CMOS SPAD with a high detection efficiency at wavelengths longer than 600 nm have been demonstrated for example by Webster [22] and Takai [23] (e.g. 30% at 800 nm), but the photo-generated carriers are collected from extended neutral (or low-field) regions, resulting in a slow temporal response. On the contrary, when the collection of the carriers is limited to well-defined depleted regions, the detection efficiency in the red and near-infrared spectrum drops to significantly lower values [24, 25] (e.g. 12% at 800 nm). These issues can be mainly traced back to the absence of a thick (e.g. 10 µm) quasi-intrinsic layer insulated from the substrate.

Alternative to CMOS-SPADs are the so-called custom-technology SPADs. In this case the detector is fabricated by using a process flow specifically tailored for SPAD manufacturing. The advantage of adopting this approach lies in the possibility of completely customize the fabrication process to obtain the desired detector structure. Custom-technology SPADs with excellent detection efficiency in the red and near infrared region (e.g. 60% @ 800 nm) have been available, also commercially [26, 27], for many years [28]. They are usually referred to as *thick SPAD*. Unfortunately, these detectors provide a poor timing jitter, of some hundreds of picoseconds [29]. Moreover, their non-planar structure is inherently not compatible with the fabrication of arrays. On the contrary, *thin SPADs* provide a timing jitter as low as 30 ps, but their detection efficiency at the longer wavelengths is somehow limited (e.g. 17% @ 800 nm) [30].

To overcome the limitations of custom-technology thick and thin SPADs, a few years ago we introduced the so-called red-enhanced SPAD (RE-SPAD) [31]. This detector achieves a remarkable detection efficiency in the red and near infrared region (e.g. 40% @ 800 nm) while preserving a good temporal response with a timing jitter as low as 90 ps FWHM. However, in its first generation, RE-SPAD did not allow for the fabrication of arrays. On the one hand, this paved the way to the successful exploitation of these detectors in a large number of applications that requires single-pixel detectors with high Photon Detection Efficiency (PDE) and low timing jitter (e.g. [32-34]); on the other hand, there is an always growing number of applications that, beside requiring high PDE and low timing jitter, would greatly benefit from the availability of SPAD arrays, even with only a few tens' of pixels. Among them are for example high-throughput single-molecule spectroscopy [2, 35], non-line-of-sight imaging [35], super-resolution microscopy [37], and time-domain diffuse correlation spectroscopy [38].

To address the needs of these applications we developed a new generation of RE-SPADs. In particular, in this paper we will present the design, the fabrication and the experimental characterization of a second generation of red-enhanced technology specifically conceived for the fabrication of SPAD-arrays. In section 2 we will briefly review the first generation red-enhanced technology and its limitations; in section 3 we will discuss the requirements for fabricating arrays of red-enhanced SPADs; in sections 4, 5, and 6 we will present respectively the new device structure, its fabrication process, and we will discuss its advantages and potential limitations; in sections 7, 8, and 9 we will provide a thorough experimental characterization of the new technology that spans from wafer-level measurements to array properties; finally, in section 10 we will draw some conclusions.

## 2. Thin and Red-Enhanced SPAD structure

In this section we will briefly introduce the structure of thin and red-enhanced SPADs. In particular, we will show that double epitaxial thin SPADs present numerous features that make these devices especially suited for the fabrication of low timing jitter arrays. We will discuss briefly these features and we will show why they lose their effectiveness in first-generation red-enhanced SPADs.

*Thin SPAD*

Fig. 1.a shows the cross-section of a typical double-epitaxial thin SPAD [39, 40]. The core of the detector is in the central part of the figure where the *shallow n* region represents the cathode and the active area is laterally limited to the *enrichment*. The latter is indeed a p+ region that lowers the breakdown voltage compared to the surrounding area, thus confining the avalanche process in the central part of the detector and thus preventing the onset of edge breakdown (virtual guard-ring effect). The *enrichment* plays a key role also in optimizing the detector performance because, by changing its doping profile, it is possible to fine-tune the shape of the electric field ([41]). The epitaxial *buried layer* and the *sinker* provide a low-resistance path for the current that flows through the active area and that is collected at the lateral *anode* contact. Very important for the implementation of SPAD arrays are the n-type *substrate* and the deeply diffused n-type *isolation*. They form a single n-type region that can be used to electrically isolate two adjacent pixels, so that they can be operated independently. As the electrical isolation is obtained by reverse biasing the pn junctions between the anodes and the substrate, we usually refer to it as *junction isolation*. Incidentally, the n-type *substrate* prevents also the detection of carriers photo-generated into the substrate, thus avoiding very slow components in the detector temporal response [40].

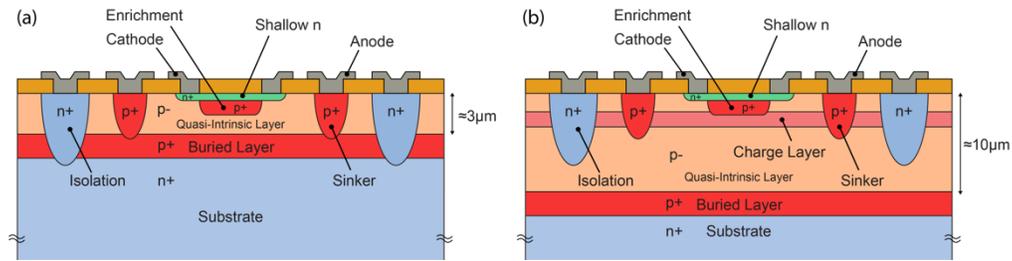

Fig. 1. Cross section of a thin, double epitaxial SPAD (a) and of a 1st-generation RE-SPAD (b).

The photons emitted during the avalanche process can trigger a correlated event when absorbed in a nearby SPAD, resulting in the so-called optical crosstalk. The *isolation* region plays also a fundamental role in preventing the direct component of the optical crosstalk because the high doping level of the *isolation* makes it very efficient in absorbing the photons that travel along a straight line connecting two detectors [42].

*First-generation red-enhanced SPAD*

In thin SPADs, red and near-infrared detection efficiency is limited by the probability of absorbing a photon within the relatively thin p- epitaxial layer. To overcome this limitation, we introduced the red-enhanced SPAD. Fig. 1.b shows a cross section of a first-generation red-enhanced SPAD. The device structure can be regarded as an evolution of the standard thin SPAD, in which a thicker p- epitaxial layer (*quasi-intrinsic layer*) guarantees a larger photon absorption efficiency for the photons in the red and near infrared portion of the spectrum. Actually, we demonstrated that the optimization of the electric field profile requires that a suitable amount of boron be inserted at a depth of about 2 μm from the device surface [43]. A simple way of performing this operation is to dynamically change the flow of the boron-dopant gas during the growth of the epitaxial layers. This solution allows the introduction of the boron at any desired depth from the surface but results in a layer that extends across the whole wafer and that is indicated as *charge layer* in Fig. 1.b. An obvious drawback of the presence of the *charge layer* beneath the edge of the *shallow n* is the reduction of the edge-breakdown voltage $V_{BD,EDGE}$. To cope with this effect, the red-enhanced structure may also include some guard rings (two for the case represented in Fig. 2) that allow the biasing of the SPAD active junction at a voltage larger than $V_{BD,EDGE}$. Indeed, the voltage $V_{KA}$ that can be applied between the SPAD anode and cathode, without exceeding the edge breakdown at the outer ring, is limited to:

$$V_{KA} < n \cdot \Delta V_{GR,MAX} + V_{BD,EDGE} \tag{1}$$

where n is the number of guard rings and $\Delta V_{GR,MAX}$ is the maximum voltage that can be applied between two adjacent guard rings [44].

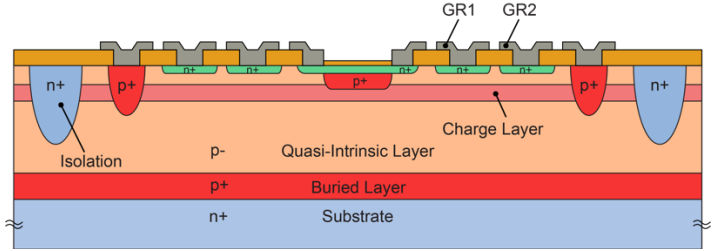

Fig. 2. Cross section of a first-generation red-enhanced SPAD with guard ring structures (GR1 and GR2). Guard rings allow to operate the detector at a voltage larger than the edge breakdown.

It's worth noting that guard rings are not needed if the sum of the breakdown voltage $V_{BD}$ and of the overvoltage $V_{OV}$ at which the detector is operated does not exceed $V_{BD,EDGE}$. However, given a certain $V_{BD,EDGE}$, guard rings may provide the flexibility needed to operate the detector at larger $V_{OV}$ and/or to design SPADs with a larger $V_{BD}$ (potentially useful to optimize other device parameters).

Compared to the thin SPAD, the red-enhanced structure represented in Fig. 1.b has two obvious limitations: on the one hand, the isolation region does not reach through the substrate, causing the loss of isolation between the anodes of multiple pixels fabricated on the same chip; on the other hand, the sinker does not get in contact with the buried layer, leading to a remarkable increase of the SPAD series resistance. As will be discussed in the following section, these two issues seriously limit the possibility of fabricating arrays of SPADs using the first-generation red-enhanced structure.

## 3. Arrays requirements

In this section we will examine which are the features that the RE-SPAD structure should have in order to enable the fabrication of low timing-jitter arrays and we will conclude that a second-generation structure is needed.

We previously demonstrated that attaining very low jitter, down to a few tens of picoseconds, requires the detection of the avalanche when it is still confined in a small area around the seeding point [45]. This task is usually accomplished by injecting the avalanche current in a small-value resistor and by detecting the increase in the voltage-drop by means of a fast comparator operated at a few-mV threshold [45]. Alternatively, the avalanche current can be injected in a high-bandwidth transimpedance amplifier and the corresponding signal can be detected again with a fast voltage-comparator [46, 47]. Whatever is the solution adopted, we will refer to these electronics as the Low-Jitter Avalanche-Readout Circuit (LJARC). At the same time, the voltage applied to the SPAD must be lowered below the breakdown value, to quench the avalanche and to allow for the release of trapped carriers, and must eventually be restored to the initial value for a new photon-detection to happen. The sequence of these operations is usually performed by the so-called Active Quenching Circuit (AQC) [14, 48].

Typically, the AQC and the LJARC are distinct circuits, which are connected at the two terminals of the detector, as in Fig. 3.a. This architecture requires a full isolation of the array, i.e. the pixels can have neither the cathodes nor the anodes in common. While the electrical isolation between the cathodes is automatically obtained thanks to the p-type epitaxial layer, the isolation between the anodes must be implemented using additional structures that surround each pixel. Despite this additional complexity, electrical isolation between anodes is strongly advisable also when a single circuit must be connected to each pixel of the array (e.g. if the LJARC is not needed or if it has been integrated within the AQC [49]). Indeed, the application of fast and large voltage-transients at the cathode, as those needed for quenching and resetting the SPAD, is complicated by the possible presence of the guard rings, which must be biased at an almost fixed voltage with respect to the cathode itself. Therefore, either with or without a LJARC, it is mandatory or highly preferable to connect the AQC at the anode. Clearly, the first-generation red-enhanced structure does not allow for this and should be modified to recover a full isolation between the pixels.

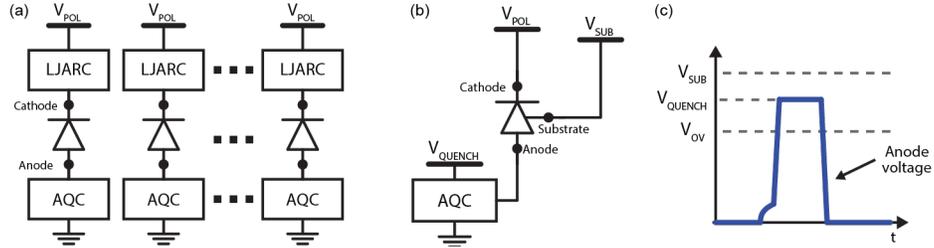

Fig. 3. (a) Typical architecture of a low-jitter SPAD-array, with the Active Quenching Circuit (AQC) and the Low-Jitter Avalanche-Readout Circuit (LJARC) connected at the two terminals of the detector. (b, c) Sketches outlining respectively the different voltages involved in the operation of the SPAD connected to the AQC, and the waveform at the anode.

The choice of connecting the AQC at the SPAD anode has some important implications on the requirements that the breakdown voltage of the anode-substrate junction must comply with. Fig. 3.b shows the typical connection of a pixel to the AQC and voltage supplies. When waiting for a photon arrival, the AQC maintains the anode voltage at 0V and the voltage applied to the SPAD is $V_{POL} = V_{BD} + V_{OV}$. Upon the detection of an avalanche, the AQC raises the anode voltage to $V_{QUENCH}$ (Fig. 3.c), which must be chosen such that $V_{QUENCH} > V_{OV}$ to ensure the proper quenching of the avalanche. On the other hand, to avoid that the substrate-anode junction becomes forward biased during this phase, the substrate must be biased at a voltage $V_{SUB} > V_{QUENCH}$. But $V_{SUB}$ is also the voltage applied at the substrate-anode junction during the photon-waiting phase, as in this phase the anode grounded by the AQC. So $V_{SUB}$ cannot exceed the breakdown voltage $V_{BD,\ SUB-A}$ of the substrate-anode junction. Combining all the requirements together, we obtain:

$$V_{OV} < V_{QUENCH} < V_{SUB} < V_{BD,SUB-A} \qquad (2)$$

The equation (2) shows how the breakdown voltage $V_{BD,\,SUB-A}$ limits the maximum overvoltage $V_{OV}$ at which the SPAD can be operated. Therefore, to be able to use the detector at the desired overvoltage, $V_{BD,\,SUB-A}$ should be sufficiently high. In order to quantify this requirement, we can notice that red-enhanced SPADs attain the best trade-off between dark count rate and PDE at relatively high overvoltages due to the thicker depletion layer compared to thin SPAD; a typical operating value is $V_{OV}=20V$ [31]. On the other hand, every inequality in equation (2) must be satisfied with some safety margin to account for possible tolerances and/or overshoots during fast transients. Therefore, the minimum acceptable value for the substrate breakdown-voltage is $V_{BD,\,SUB-A}|_{MIN}= 25V \div 30V$, and higher values are certainly preferable because they provide the freedom of operating the detector at higher overvoltages. This can be useful not only to increase further the PDE on the current red-enhanced detectors (at the expenses of a higher DCR), but also to have the possibility of designing red-enhanced SPADs with a thicker epitaxial layer (at the expenses of a higher timing-jitter), which would require a correspondingly higher $V_{OV}$. As first-generation RE-SPADs have $V_{BD,\,SUB-A} \cong 25V$, which is the very minimum acceptable value, some improvement is needed also on this side.

Another requirement a SPAD must comply with for attaining a low timing-jitter, either as a single-pixel or arranged in an array, is to have a low series resistance. Indeed, we previously demonstrated that the resistance, that is due in part to the resistive path the current travels through the device and in part to the space charge effects, plays a key role in determining the avalanche propagation velocity and the associated jitter [50]. In particular, a lower resistance results in a faster avalanche propagation with a correspondingly lower timing jitter. Obviously, the structure of Fig. 1.b is not optimal from this point of view and improvements are highly needed.

Finally, it's worth noting that the limitations of the structure of Fig. 1.b, although not ideal, do not prevent the operation of the detector in a single-pixel configuration. Indeed, electrical isolation is not needed with a single-pixel, series resistance can be partially reduced be enlarging considerably the sinker region, and the low value of $V_{BD,\,SUB-A}$ can be dealt with by operating the SPAD with the substrate in a semi-floating configuration [44]. However, these solutions are not applicable to arrays, where the electrical isolation between pixels is mandatory, a larger sinker would compromise the fill factor, and the substrate must be biased at a fixed voltage to avoid disturbances between pixels.

Therefore, the structure of Fig. 1.b should be regarded as a temporary and simple solution that allowed us to demonstrate the properties of red-enhanced SPADs in terms of PDE and timing jitter, in view of a more advanced structure, which is fully compatible with the fabrication of arrays.

### 4. Second generation Red-Enhanced structure

To satisfy the requirements discussed in the previous section, we introduced a new detector structure, which is depicted in Fig. 4. Compared to the previous version of Fig. 1.b, there are two main modifications: on the one hand, we replaced the *n-type isolation* region with a *deep trench*; on the other hand, we added a lightly doped n-type layer, which we will call *interposed layer*, in between the buried layer and the substrate. For the sake of clarity, the guard rings are not shown in Fig. 4, although they may still be added to this structure, if needed.

*Interposed Layer*

The role of the *interposed layer* is to increase the breakdown voltage at the anode-substrate junction. Indeed, as the layer is very lightly doped (dopant concentration $< 10^{14}$ cm$^{-3}$), the interposed layer is fully depleted and the depletion region extends from the buried layer to the substrate. To a first approximation, the electric field $E_{IL}$ in the depletion layer is constant and can be calculated as $E_{IL}= V_{SUB-A} / t_{IL}$ where $t_{IL}$ is the thickness of the *interposed layer* and $V_{SUB-A}$ is the voltage applied at the anode-substrate junction. As the breakdown happens for an electric field in the order of $E_{BD} = 10^5$ V/cm or, equivalently, $E_{BD} = 10$ V/µm, a thickness $t_{IL}$ of some µm is sufficient to reach a breakdown voltage of many tens of volts, in accordance with our requirements. Notably, the design (and fabrication) of the *interposed layer* is not critical, as we only need that the breakdown voltage is higher than a certain threshold. Therefore, any doping level low enough and any thickness large enough meet the requirements.

*Deep Trenches*

The main role of the *deep trenches* is to provide the full isolation between the pixels of an array. The idea is to place an insulating layer in between the anodes; if such a layer reaches through the n+ substrate, any conductive path between the anodes is effectively suppressed. A practical way of implementing this solution is to etch a trench deep into the silicon and to cover its sidewalls with silicon dioxide. The trenches represented in Fig. 4 are also refilled with polysilicon for reasons that will be discussed later.

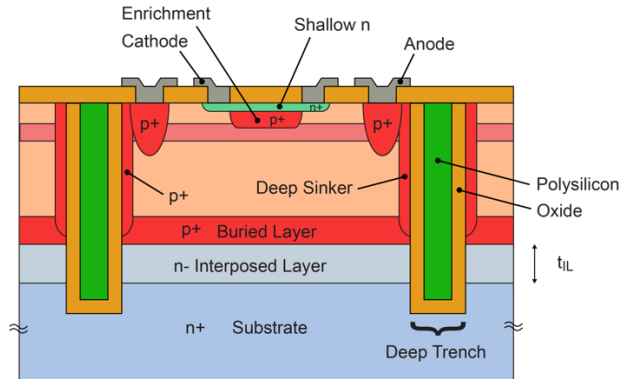

Fig. 4. Cross section of a second-generation red-enhanced SPAD. The n- interposed layer increases the anode-substrate breakdown voltage; the deep trenches and the deep sinker provide respectively full electrical insulation between pixels and low series resistance.

Considering two adjacent pixels, the maximum voltage drop across the trench in between them is $V_{QUENCH}$. This value is reached when one of the anodes is kept to ground (photon-waiting phase) and the other one is kept to $V_{QUENCH}$ (hold-off phase) by the corresponding AQCs. As quenching the avalanche requires only that $V_{QUENCH} > V_{OV}$, the trenches need to withstand a voltage drop at most of 30V - 40V. These values can be easily obtained with the proposed solution because the dielectric strength of silicon dioxide is in the order of 1 V/nm. Therefore, an $SiO_2$ layer with a thickness of a few tens of nanometers is sufficient to the purpose.

Deep trenches can be exploited also to solve the high-resistance issue, which is due to the separation between the sinker and the buried layer. Indeed, during trench fabrication it is possible to implant some boron in the trench sidewalls to attain a low resistivity path along the trench side. This path, indicated as *deep sinker* in Fig. 4, allows us to recover a good electrical contact between the sinker and the buried layer. It's worth noting that the *deep sinker* must stop into the *buried layer*, otherwise different problems can occur as outlined Fig. 5. For example, if the *deep sinker* goes all-around the trench (Fig. 5.b) the isolation between the anodes is lost. Similarly, if the deep sinker terminates into the interposed layer or into the substrate (Fig. 5.c) the effectiveness of the interposed layer is (partially) compromised. In fact, the distance between the p+ region of the anode and the n+ region of the substrate becomes smaller than $t_{IL}$ leading to a reduction of the corresponding breakdown voltage (curvature effect could make the situation even worse).

Deep trenches may also be used to suppress the direct optical crosstalk. Indeed, the photons emitted by a first SPAD during the avalanche, and travelling straight toward a nearby detector, can be absorbed by the highly doped polysilicon or reflected/scattered at the trench sidewalls.

However, the effectiveness of this solution is difficult to assess during the design and must therefore be verified experimentally.

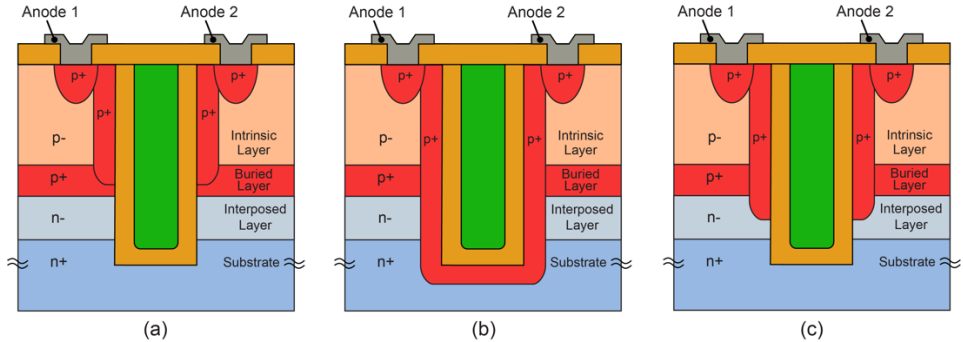

Fig. 5. In order not to compromise the effectiveness of the deep trench and of the interposed layer, the deep sinker must stop into the buried layer (a) rather than surround the trench (b) or reach through the interposer layer (c).

## 5. Fabrication process

In the previous section we presented a new structure that addresses the limitations of the first-generation red-enhanced technology in terms of development of low-jitter arrays. In this section we will discuss the modification to the fabrication process that are needed to implement the two key features of the new structure, namely the *interposed layer* and the *deep trenches*.

The fabrication of the *interposed layer* is trivial. It can be grown epitaxially right before the epitaxial growth of the buried layer and of the p- layer. In practice, the three layers are grown in a single epitaxial process during which the flow of the dopant gases is dynamically changed to meet the desired doping profile and thicknesses.

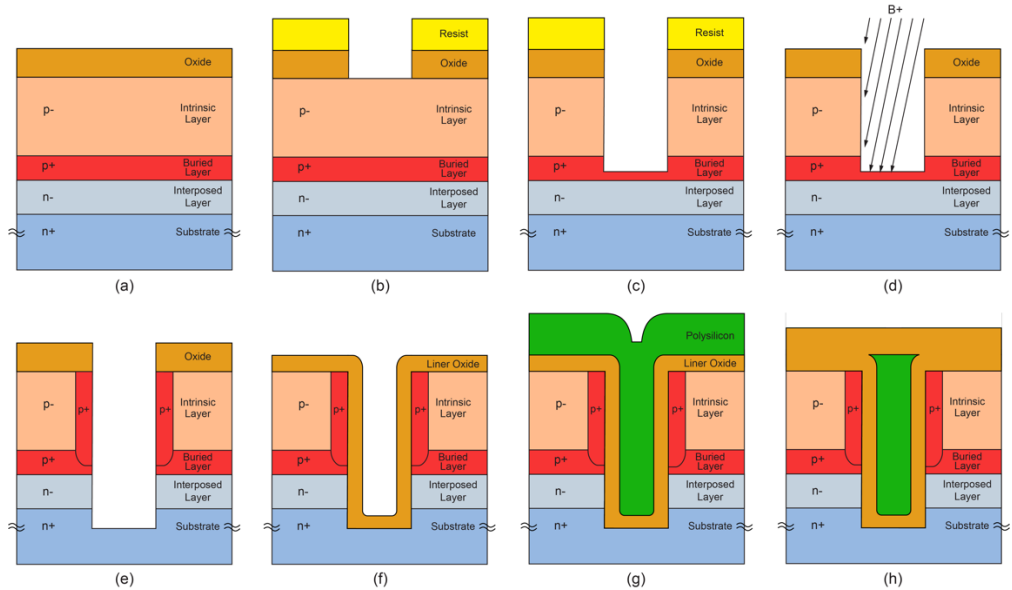

Fig. 6. Deep trenches fabrication process. The epitaxial wafer is covered with a thick oxide to serve as a hard mask (a) and is patterned through standard optical lithography (b). The silicon is initially etched down to the buried layer (c) and some boron is implanted into the trench sidewalls to form the deep sinker (d). The silicon is then etched down to the substrate (e) and the damaged hard-mask oxide is replaced with a new oxide layer that serves as insulating material (f). Finally, the trench is refilled with polysilicon (g), and the wafer surface is planarized and covered with an additional oxide layer (h).

The main steps needed for manufacturing the *deep trenches* are illustrated in Fig. 6. After the growth of the epitaxial layers, the wafer is covered with a thick layer (≈ 1 μm) of silicon dioxide to serve as a hard mask during trench etching (Fig. 6.a). The trench pattern is transferred into the hard mask by standard optical lithography followed by $SiO_2$ reactive ion etching (Fig. 6.b). The exposed silicon is then anisotropically etched down to the buried layer by means of a deep reactive ion etching (DRIE) process (Fig. 6.c). After stripping the photoresist from the wafer surface, the deep sinker if fabricated by ion implantation followed by thermal drive in. In particular, a high boron dose is implanted into the trenches sidewalls by tilting the wafer with respect to the implantation beam (Fig. 6.d). To ensure the formation of the deep sinker on every sidewall, independently from the orientation of the trench with respect to the wafer, the implantation process is repeated four times, rotating the wafer of 90 degrees in between each iteration. A second step of DRIE is then performed to etch the trench down to the n+ substrate (Fig. 6.e). The splitting of the DRIE process in two different steps, separated by the sidewalls boron implantation, allowed us to stop the deep sinker into the buried layer, which is necessary for the reasons discussed in the previous section. Terminated the etching of the trenches, the damaged oxide of the hard mask is removed and replaced with a new layer of silicon dioxide (*liner oxide*), which acts also as an insulator for the trench sidewalls (Fig. 6.f). Deep trenches are then refilled with phosphorus-doped polysilicon (Fig. 6.g). Refilling is necessary to obtain a flat surface that allows the routing of the interconnections (metal lines) across the trenches. The polysilicon has been chosen as a refilling material because it can be deposited conformally and because it contributes to the absorption of the photons responsible for the optical crosstalk, especially if heavily doped. Moreover, polysilicon has a thermal expansion coefficient similar to monocrystal silicon therefore it contributes to reduce the stress introduced by the trenches during high-temperature processes. The top polysilicon layer is then removed and the wafer surface simultaneously planarized by using a Chemical Mechanical Polishing (CMP) process. The deposition of an additional layer of $SiO_2$ allows us to insulate and protect the polysilicon during the following processing steps (Fig. 6.h). From here on the fabrication of the red-enhanced SPAD can proceed exactly as in the first-generation devices.

## 6. Advantages and limitations

The device structure and the related fabrication process we described respectively in Section 4 and Section 5 have some advantages and potential issues. This section is dedicated to their discussion.

*Easiness of integration in the SPAD process flow*

Generally speaking, when an additional step is introduced in a fabrication process, it may have a significant impact both on the structures that have already been fabricated and on those that are still to be manufactured. The former effect happens because of the additional thermal treatments associated with the new step, which may lead to a diffusion of the dopants already present in the wafer; the latter effect happens because the layers that are removed or added during the new step may change the wafer topography. For this reason, adding some steps in a consolidated process flow is usually critical and may require the redesign of a significant part of the process to compensate for the effects introduced by the new steps.

One of the advantages of the structure we adopted, is the easiness of its integration in the overall SPAD fabrication process. This happens thanks to three key properties: the trench fabrication is entirely performed before the manufacturing of the SPAD core; the process flow has a small impact on the overall thermal budget; and the trenches do not impact on the wafer topography. More in detail, the fact that the trenches are fabricated right after the completion of the epitaxial growth means that all the thermal treatments associated with their fabrication are performed before the manufacturing of the SPAD active area. Therefore, the doping profile in the active area is not affected by the trench manufacturing and a redesign of this critical region is not needed. Moreover, the last two steps of the trenches' fabrication, illustrated in Fig.

6.h, leave the wafer surface completely flat and covered with a thick oxide, as it would happen in absence of trenches. Therefore, the core of the SPAD fabrication process can be executed neglecting the presence of the trenches.

By contrast, the thermal treatments performed during trench fabrication may lead to a diffusion of the dopants already present in the epitaxial layers. Especially detrimental for the SPAD performance is the diffusion of the boron in the buried layer because the lifetime of the diffusion tail (i.e. the slow component of the temporal response) increases quadratically with the thickness of the buried layer itself [40]. For this reason, it is especially important that trench fabrication has a negligible effect on the overall thermal budget. In evaluating this aspect, the steps that must be considered are those that are performed at high temperature, i.e. hard mask creation, deep sinker annealing, liner oxide growth, and polysilicon deposition. As the liner oxide remains on top of the wafer at the end of the trench fabrication, it can replace the field oxide that is usually grown at the beginning of the SPAD fabrication, so its thermal budget is compensated. The hard mask can be largely deposited at low temperature, so its impact is negligible. Similarly, the thermal budget associated with the annealing of the deep sinker and with the deposition of the polysilicon is small compared to other steps in SPAD fabrication.

In conclusion, the deep trenches can be regarded as a process module that can be optionally introduced in the usual SPAD fabrication process with small impact.

*Compactness*

Another advantage of the proposed structure is its compactness, which is of the utmost importance in view of the fabrication of dense arrays of red-enhanced SPADs. Indeed, it is possible to manufacture trenches that are deep and at the same time narrow, thanks to the availability of highly anisotropic silicon-etching processes (e.g. the so-called Bosch process [51]). Aspect ratios (depth-to-width ratio) of 10:1 or 20:1 can be easily achieved and allowed us to fabricate 3 µm wide trenches (smaller values can also be attained). This represents an interesting result not only for red-enhanced SPAD, but also a significant improvement compared to thin SPAD technology where the *isolation region* is used instead of the deep trenches. In fact, the minimum width of the isolation region is around 10-12 µm, considering that the dopant must diffuse vertically for at least 6-7 µm (to reach the substrate) and that, at the same time, it diffuses laterally for about 70% as much (in both directions). In addition, a minimum distance of a few microns between the sinker and the isolation must be preserved to avoid the breakdown of the corresponding pn junction. On the contrary, the sinker can be placed as close to trench as the fabrication tolerances allow.

*Parasitic capacitances reduction*

The introduction of the interposed layer, in addition to increase the breakdown voltage, reduces the parasitic capacitance of the anode-substrate junction. As this capacitance must be discharged and recharged by the AQC every time an avalanche is detected, its reduction may result in a faster quench and reset time along with a reduced dynamic power dissipation.

*Effect on dark count rate*

A well-known problem associated with the proposed structure is the stress that develops in proximity of the deep trenches. The main causes are: (1) the difference in thermal expansion coefficients of silicon and silicon dioxide that results in a considerable stress during high temperature treatments of the wafers; (2) the progressive shrinkage of the polysilicon refill due to the grains growth that happens during high temperature processes [52]. Above a certain threshold, the stress can generate structural defects in the silicon crystal, which can have a dramatic impact on the DCR if they reach the SPAD active area. In particular, it has been clearly shown that dislocations can originate from the bottom of the trench and can propagate till the surface of the wafer where they are visible as slip lines [53].

Despite this potential issue, two aspects suggest that deep trenches can be effectively used for SPAD fabrication: on the one hand, it is possible to avoid or strongly limit the formation of defects with a suitable design of the trench layout and fabrication process (i.e. trench width, oxide thickness, etc.); on the other hand, dislocations propagates along the {1 1 1} crystallographic plane, with an angle of about 35 degrees with respect to the trench axis [53, 54]. Therefore, the defective region, if present, is confined in a small region around the trenches and any negative effect on the DCR can be prevented by increasing the distance between the trench and the active area. The effectiveness of these solutions will be verified experimentally.

*Other potential advantages*

The new structure provides also opportunities to further improve detector performance. For example, once confirmed that deep trenches refilled with highly doped polysilicon are sufficient to suppress direct optical crosstalk (see section 9), it will be possible to completely remove the process for the formation of the isolation region. As the phosphorus must be diffused down to the substrate, this process has a very high thermal budget. Removing this process will allow us to attain a thinner buried layer resulting in an improved diffusion tail.

## 7. Wafer-level measurements

In order to verify the effectiveness of the solution proposed in the previous sections, second-generation red-enhanced SPADs have been fabricated by authors at the Cornell NanoScale Science and Technology Facility (CNF), Cornell University, Ithaca (USA). In sections 7, 8, and 9 we will report a thorough characterization of these devices.

*Dark count rate distribution*

As pointed out in the previous section, one of the main unknowns of introducing the deep trenches is their effect on the DCR. For its assessment, we fabricated devices both with first-generation and second-generation structure and compared their DCR. To ensure a fair comparison, the two types of detectors have been fabricated starting from identical substrates on top of which we grew the same epitaxial layers, including the interposed layer. On the wafer W11 we manufactured the deep trenches, while on the wafer W10 we grew only a suitable field oxide. Starting from this point, the two wafers have undergone exactly the same process flow. In particular, all the steps for the creation of the *isolation region*, which is needed on the wafer W10, have been performed also on the wafer W11. This guarantees exactly the same thermal budget on the two wafers, and therefore the same doping profile for a fair comparison of their performance.

On each wafer, we performed a split of the enrichment dose. In practice, we repeated the implantation process multiple times, each time masking selected regions of the wafer. We used four different boron doses, namely D1 – D4, chosen to span a large set of breakdown values, respectively of about: 110 V, 60 V, 35 V, and 25 V.

To acquire a statistically significant amount of data, the DCRs have been measured at wafer level by using a semi-automated probe-station. A printed circuit board mounted on one micromanipulator allowed us to connect an AQC to one terminal of the SPAD; another micromanipulator allowed us to apply to the other terminal the voltage needed to bias the SPAD at the desired $V_{OV}$.

Due to the lack of isolation, on the wafer W10 it is not possible to operate the AQC connected to the anode (which is common to all the SPADs of the wafer). On the contrary, this would be possible on the wafer W11 thanks to the deep trenches. However, in the interest of a fair comparison, we decided to characterize both the wafers operating the SPADs with the AQC connected to the cathode. To this regard, an issue, which occurred during fabrication, resulted in guard rings that are shorted with the cathode. On the one hand, this greatly simplified the operation of the SPADs with the AQC connected to the cathode, as the guard rings automatically follow the voltage transients at the cathode; on the other hand, it prevented the exploitation of the guard rings to bias the detector at voltages higher than the edge breakdown, which is about 55 V for both W10 and W11 (measured on purposely developed SPADs without the enrichment region). The lack of usable guard rings prevents the operation of the SPADs with the higher breakdowns (dose D1 and D2); therefore, we focused on the characterization of the devices with a breakdown voltage of about 35 V and 25 V (dose D3 and D4).

Initially, we compared detectors without trenches against detectors in which the trenches are placed on an 800 μm side square, centered around the SPAD. This is an especially forgiving situation because the trenches are far from the active area (minimum distance 375 μm), therefore it is unlikely that the defects, that possibly originate from the trenches, propagate through the active area. Nevertheless, it is an interesting case because it would enable the fabrication of coarse arrays of SPADs, e.g. with pixel-to-pixel pitch of 500 μm, as required for example in single-molecule analysis [55, 56]. The results of this comparison are shown in Fig. 7.a, which represents the inverse cumulative distribution of the DCR. Each curve has been obtained by sorting each set of detectors in increasing order of DCR and plotting their DCR against the detector number, normalized to 100. These curves allow us to read on the x-axis the percentage of devices whose DCR is lower than the corresponding value on the y-axis; therefore, they are particularly useful to compare different sets of SPADs. Blue and red curves correspond to single-pixel detectors respectively belonging to wafers W10 (without trenches) and W11 (with trenches). In both cases the detectors have an active area diameter of 50 μm and the enrichment dose is D3. The measurements have been performed at a temperature T= 23 °C, set by a temperature-controlled chuck, and at an overvoltage $V_{OV}$= 15V. The fact that the two DCR distributions are almost identical allows us to conclude that the trench-fabrication process is clean (i.e. it does not introduce contaminants up to a significant level) and that the defects possibly generated by the trenches do not propagate over long distances. This confirms that deep trenches are a viable solution for coarse arrays of SPADs.

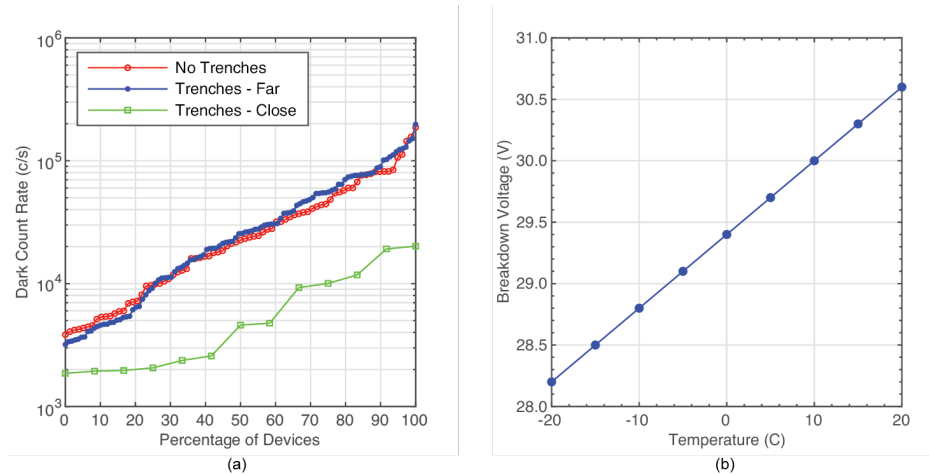

Fig. 7. (a) Distribution of the dark count rate for 50 μm diameter SPADs without trenches (red open circles), with trenches placed far from active area (blue closed circles), and close to the active area (green open squares). All measurements have been performed at an overvoltage $V_{OV}$= 15V and a temperature T= 23°C. (b) Breakdown voltage as a function of the temperature as measured on die O25.

To evaluate the perspectives of developing arrays of SPADs that are more compact, on wafer W11 we designed also detectors with trenches much closer to the active area (minimum distance 60 μm). In the following we will refer to these detectors as *compact SPADs*. The corresponding DCR distribution is represented in Fig. 7.a in green. By comparing this curve with the others, it is clear that deep trenches do not lead to an increase of the DCR even when relatively close to the active area. On the contrary, it may seem that they have a beneficial effect on the DCR. However, we cannot draw a final conclusion on this point because, on the one hand, it has not been possible to measure these devices on a temperature-controlled chuck and, on the other hand, their breakdown is slightly higher compared to the others. These two facts might be responsible of the lower DCR; this aspect will be investigated more in detail in the future.

## 8. Packaged devices measurements

To perform a complete characterization of the key SPAD parameters, we diced part of the wafers and packaged the corresponding chips in some TO headers with a glass-window cap. The use of packaged devices allowed us to operate these detectors in the setups purposely developed to measure each parameter.

All the DCR, PDE, and afterpulsing measurements reported in the following subsections have been performed on the same SPAD, namely O25, to give a consistent picture of the performance of one of these detectors. However, each parameter has been measured on multiple chips to gain a basic understanding of its uniformity.

The SPAD O25 belongs to the wafer W11 and has an active diameter of 50 μm. It has been selected in the first 10% of the DCR distribution shown in Fig. 7.a. A low breakdown voltage $V_{BD}$= 30V allows for a complete characterization using overvoltage values as high as 20 V without the help of the guard rings. To perform measurements at controlled temperatures down to -20 °C, the SPAD has been mounted on a Peltier stage inside a TO header. When required, the Peltier stage has been driven by an external instrument (SE5020 by Marlow Industries), which implements a closed-loop temperature-control. The temperature value is obtained by reading a thermistor placed in good thermal contact with the detector chip. In all the measurements the SPAD has been operated connected to an external AQC board.

*Breakdown voltage vs temperature*

As a preliminary step of the characterization we measured the temperature dependence of the breakdown voltage. This information is needed as we usually want to perform measurements at a fixed overvoltage, and so we want to compensate for the breakdown variation with the temperature. Fig. 7.b reports the results obtained by changing the temperature from -20 °C to 20 °C in step of 5 °C and by measuring at each temperature the breakdown voltage with a semiconductor parameter analyzer (Agilent 4155B). As expected, the curve is linear; the extracted temperature coefficient is 60 mV/°C. This is value is larger than in thin SPADs [40] owing to the thicker depleted layer.

*Dark count rate*

Fig. 8.a reports the DCR as a function of the temperature for the device O25. The measurements have been performed at a constant overvoltage $V_{OV}$= 20V. It is possible to observe the typical exponential dependence of the DCR on the temperature [41], at least at the higher temperatures. A slight change of the slope at the lower temperatures suggests that the DCR in this region becomes dominated by the band-to-band tunneling effect [41]. This is not surprising given the low breakdown voltage of the detector, which corresponds to a high peak-value of the electric field.

Because of the field-enhanced generation mechanisms [57], we think that the high peak of the electric field is in part responsible also for the relatively high DCRs observed in Fig. 7.a. Unfortunately, the issue with the guard rings did not allow us to use the detectors with a higher breakdown voltage to investigate further the effect of the electric field on the DCR.

Fig. 8.b reports the DCR as function of the overvoltage, measured at a temperature T= 20°C (red curve). For comparison we reported also the corresponding curve (blue line) for a typical thin SPAD. This figure will be discussed in the next subsections, in conjunction with the PDE results.

*Photon detection efficiency*

The PDE has been measured on a typical setup based on a monochromator (Oriel SpectraLuminator 69050) and on an integrating sphere. The monochromator generates a 10 nm bandwidth beam of light whose center wavelength can be swept between 400 and 1000 nm. The light is fed into an integrating sphere to obtain a uniform illumination on a centimeter-size area in front of the primary output port of the sphere. A preliminary calibration allows us to calculate in real time the power density at the primary output port by reading an auxiliary photodiode on the secondary output port. By placing the detector under investigation in front of the primary output port it is possible to measure the PDE as the ratio between the number of photons detected by, and the number of photons impinging on the SPAD. More details on the setup and on the calculation of PDE value can be found for example in [58].

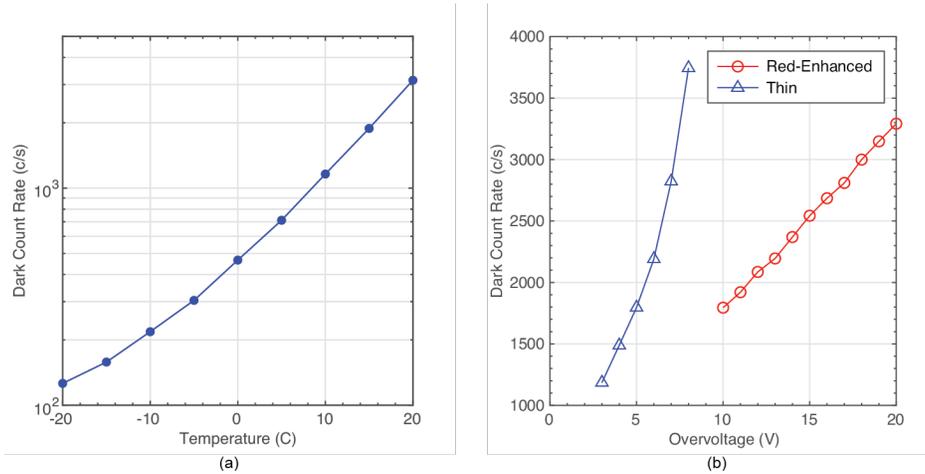

Fig. 8. (a) Dark count rate versus temperature measured on die O25 at $V_{OV}$ = 20V. (b) Dark count rate versus overvoltage for red-enhanced SPAD O25 and for a typical thin SPAD. Measurements have been performed at temperature T = 20°C.

Fig. 9.a reports the PDE, as a function of the wavelength, measured on the SPAD O25 biased at three different overvoltages: $V_{OV}$= 10 V, 15 V, and 20 V. The PDE increases remarkably with $V_{OV}$ in the considered range. This is a confirmation that red-enhanced SPAD must be operated with an overvoltage higher compared to thin SPAD. Indeed, we previously demonstrated [31] that, given a shape of the multiplication field, the overvoltage to be applied to reach a certain avalanche probability scales with the overall thickness $t_D$ of the depleted layer. In fact, a change $\Delta V$ in bias voltage determines a variation of the electric field $\Delta E = \Delta V/t_D$. Therefore, the same change $\Delta E$, and so the same change in the avalanche probability, is obtained in different devices by using an overvoltage proportional to the thickness of their depleted regions.

The need for a higher value of $V_{OV}$ does not necessarily implies a higher DCR compared to thin SPADs. Indeed, also the DCR, which is enhanced by the electric field, has a lower dependence on the overvoltage in RE-SPAD compared to thin SPAD. This happens for the same reason outlined above and can be experimentally observed in Fig. 8.b.

Although the main target of the second-generation RE-SPAD was to make the structure compatible with the fabrication of arrays, we took the chance also to adjust and refine the doping profile in the active region. For this reason, the results reported in Fig. 9.a represents a small, but significant improvement compared to the PDE previously obtained with first-generation RE-SPAD [31]. To compare these results with those typically attained with custom technologies, in Fig. 9.b we reported the PDE as function of the wavelength for: a RE-SPAD (in red), a thin SPAD (in blue), and a thick SPAD (in green). The former is again the RE-SPAD

O25, biased at $V_{OV}$= 20V. The second is a classical double epitaxial thin SPAD, developed at Politecnico di Milano [40] and commercialized by MPD Micro Photon Devices [59]; the curve reported has been obtained at the usual operating overvoltage $V_{OV}$= 5V. The latter is a typical thick SPAD, the SLiK by Perkin Elmer Optoelectronics (now Excelitas Technologies). The PDE of this detector has been measured operating the SPAD in the original SPCM-AQR module with the original (unknown) biasing conditions. The results obtained are in agreement with the values reported in [60].

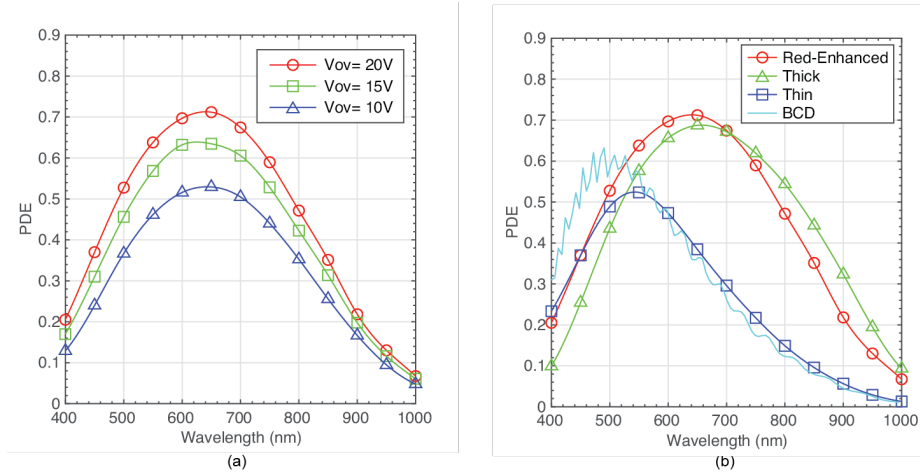

Fig. 9. (a) Photon detection efficiency versus wavelength for the red-enhanced SPAD O25 at three different overvoltages: $V_{OV}$= 10 V, 15 V, and 20 V. (b) Photon detection efficiency of different SPADs: a red-enhanced (die O25, Vov = 20V), a typical custom thin-SPAD (Vov = 5V), a thick-SPAD from Perking Elmer Optoelectronics, and a BCD SPAD (Vov = 5V).

We can observe that the red curve and blue curve are very similar in the 400 – 550 nm range. On the one hand, this is largely expected since the thin SPAD and RE-SPAD have an almost identical structure in the top region, where blue and green photons are absorbed. On the other hand, this is also a confirmation that Vov= 20V for the RE-SPAD is equivalent to $V_{OV}$= 5V for the thin SPAD, in the sense that they provide the same avalanche probability at the edge of the space charge region [43]. Conversely, the advantage of the RE-SPAD compared to the thin SPAD is evident in the red and NIR region of the spectrum, with the PDE that has a peak of 70% at 650 nm and is still above 45% at 800 nm. The thick SPAD still presents a slightly higher PDE in the NIR region, due to the considerably thicker absorption layer, which, however, negatively affects the time jitter.

For the sake of completeness, in Fig. 9.b we reported in light blue also the PDE of a SPAD fabricated in a customized BCD technology [25], which represents the current state of the art in terms of high detection efficiency and low timing jitter for non-fully custom technologies. This detector attains the higher PDE in the 400 – 500 nm range thanks to a reverse structure in which the avalanches generated by short-wavelength photons are prevalently triggered by electrons rather than by holes.

*Afterpulsing*

The afterpulsing is a phenomenon due to the carriers that flow through the detector active area during an avalanche. If one of them gets trapped into a localized state and is released at a later time, it can trigger a secondary avalanche correlated to the first one. This phenomenon can be quantified by means of the Afterpulsing Probability (AP), defined as the probability that a primary event triggers a secondary avalanche.

Afterpulsing characterization is important not only because the afterpulsing phenomenon can be detrimental for many applications that measure temporal correlations, like Fluorescence

Correlation Spectroscopy (FCS) (see for example [61]), but also because it can provide useful insights on detector properties. In particular, measuring the AP is a mandatory step to validate the measured PDE values. Indeed, the afterpulsing phenomenon results in more than one count per detected photon; therefore, if larger than a few percent, the afterpulsing can significantly distort the measured value of the PDE.

To measure the AP, we resorted to an in-house developed time-tagger. This module allows us to record the time of arrival of every photon with a resolution of 8 ns and to download the stream of the arrival times on a PC by means of an USB interface. A purposely developed Matlab program allows us to calculate the autocorrelation function (ACF) of the recorded events.

Fig. 10.a shows the autocorrelation function measured on the SPAD O25, biased at an overvoltage $V_{OV}= 20V$ and at a controlled temperature T= 20°C. At the longer delays the autocorrelation function is flat, as expected for uncorrelated events, like the dark counts. Its value is equal to the DCR of the SPAD. The increase of the ACF at shorter delays indicates the presence of correlated events. The afterpulsing probability density, i.e. the afterpulsing probability per unit time, can be obtained by subtracting from the ACF the constant contribution given by the dark counts. By integrating the afterpulsing probability density, we obtain the overall AP, which is equal to 2% for the case reported in Fig. 10.a. Similar results have been obtained also for *compact SPAD*. On the one hand, such a low value of AP validates the PDE measurements reported in Fig. 9. On the other hand, it shows that deep trenches do not have any significant effect on the AP, also when placed close to the active area.

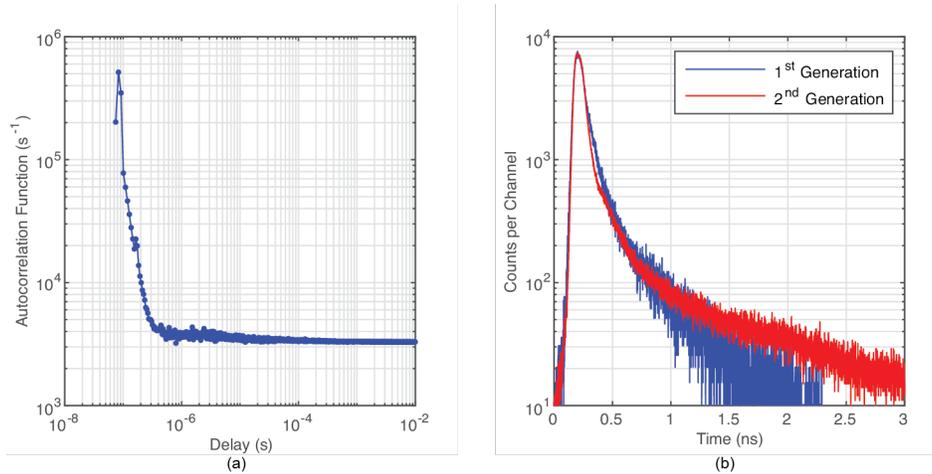

Fig. 10. (a) Autocorrelation function of counts generated in dark condition by red-enhanced SPAD O25 biased at Vov = 20V. (b) Temporal response of a first-generation (blue curve) and of a second-generation (red-curve) red-enhanced SPAD. The second-generation device attains the same timing jitter of the first-generation one, apart for a longer diffusion tail due to an accidentally thicker buried layer.

*Timing jitter*

One of the most interesting properties of red-enhanced SPADs is their ability to provide a low timing-jitter despite the high photon detection efficiency. This property has been demonstrated on the first-generation structure [31] and it must be confirmed on the second-generation structure, although we do not expect any negative effect of the deep trenches on the timing-jitter. To this aim we acquired the temporal response of the new devices by means of a conventional setup for Time-Correlated Single-Photon Counting (TCSPC) measurements.

A semiconductor laser (Antel MPL-820) generates 10-picosecond width pulses at a wavelength of 820 nm and with a repetition rate of 100 kHz. The pulses are attenuated down to the single photon level and shined on the SPAD active area. The onset of an avalanche is detected by means of the low-jitter circuit described in [62]. The distribution of the delays between the laser trigger output and the photon detection output has been acquired by means of a time-correlated single-photon counting board (Becker & Hickl SPC 830).

Fig. 10.b reports in red the results obtained on a 50 μm active diameter SPAD belonging to wafer W11. The detector has been biased at on overvoltage $V_{OV}$= 20V and has been operated at room temperature. In this curve are clearly distinguishable the sharp peak given by the photons absorbed in the depleted region and the slow diffusion tail resulting from the photons absorbed in the neutral regions above and below the depleted region itself. The peak-width is 95 ps FWHM, which confirms the good timing performance previously attained. However, the diffusion tail has a component that is especially slow. For the sake of comparison, in Fig. 10.b we reported in blue the results obtained on a red-enhanced SPAD from a previous fabrication run [31], which presents a much faster diffusion tail. The reason for this difference lies in an issue that happened during the growth of the epitaxial layers, which led to a buried layer thicker than required. As the lifetime of the diffusion tail depends quadratically on the thickness of the neutral regions, the effect is amplified in the temporal response. A doping profile analysis, carried out by using Secondary Ion Mass Spectroscopy (SIMS), confirmed the source of the problem. Moreover, the same effect is present on all the wafers processed in this run regardless of the presence of deep trenches and/or of the intermediate layer. As a consequence, we can confidently conclude that the problem is not intrinsically related to the new device structure and that it can be easily addressed in a following fabrication.

## 9. Arrays

In the previous two sections we presented a thorough characterization of single-pixel detectors aimed at demonstrating that the new structure we introduced can enable the development of arrays of red-enhanced SPAD with the same (or even better) performance we previously obtained with intrinsically single-pixel detectors. To further corroborate this idea, we actually fabricated arrays of red-enhanced SPADs. For the moment, we focused on arrays with small fill-factor, like the one needed in high-throughput single-molecule analysis [19]. We designed for example a linear array of 32 pixels having an active area diameter of 50 μm and a pitch (i.e. a center-to-center distance) of 250 μm.

After a preliminary characterization it was clear that the arrays that we developed presented two distinct issues. Firstly, the guard rings were shorted with the cathode, as it happened also on the single-pixel detectors described in the previous sections. However, in addition to this, some interconnections were also broken in between the bonding pads and the active region. While the first issue could by bypassed by using the arrays with a low breakdown voltage (i.e. with enrichment dose D3 and D4), the second one completely prevents the operation of these detectors.

By thoroughly analyzing the data collected during the fabrication and by running some specific tests, we identified the causes of the two problems. The shorts between the cathode and the guard rings are due to the unintended removal of a layer of silicon dioxide during the fabrication of the contact regions (i.e. those regions in which the aluminum gets in contact with the silicon); this happened because of wrong processing conditions adopted during the etching step. On the other hand, the broken interconnections are due to the poor adhesion of the photoresist to the aluminum. Because of this, the photoresist failed to mask the etching process and led to the unwanted removal of some aluminum along the topographic steps, where the adhesion problem happened. Although the adhesion problem was limited only to some regions of the wafers, unfortunately these regions contained all the arrays.

To solve both the issues we decided to rework the wafer W12, which is nominally identical to the wafer W11. The reprocessing essentially consisted in: removing the damaged interconnection layer; depositing a new layer of silicon dioxide to replace the one that has been erroneously etched; depositing and patterning a new aluminum layer to create new interconnections. All these processes have been performed at low temperature (< 400°C) not to alter the doping profile inside the devices.

The reprocessing actually solved the two aforementioned issues but resulted in an increase of the DCR approximately of a factor of 10. This can be ascribed to the tools and processes used during the reworking. They do not guarantee the same level of cleanliness and low-defectivity that characterizes the previous part of the process. Nevertheless, we were forced to adopt them because of the constraints posed by the low-temperature requirements and by the fact that the wafers had already been exposed to the aluminum.

Despite the higher DCR, we decided to use the arrays obtained from the reprocessing of wafer W12 to develop a compact module for high-efficiency parallel single-photon detection. The module is based on a 32x1 array of red-enhanced SPADs coupled with an array of 32x1 AQCs. It contains all the electronic circuitry needed to bias the detectors, to provide the power supply to the AQCs, to control their operation, and to deliver the photon-counting information to external systems. The SPAD array is mounted on top of a Peltier stage and can be cooled down to -15 °C to reduce the DCR; a sealed chamber, filled with dry nitrogen, prevents moisture condensation. The experimental characterization of this module confirmed the results reported in Sections 7 and 8 (apart the higher DCR) and effectively demonstrated that the second-generation structure enables the development of red-enhanced SPAD-arrays. A detailed description of the module and of its performance has been reported in [63]. In this paper we will limit the discussion to the use of the module to demonstrate the effect of the deep trenches on the optical crosstalk.

*Optical crosstalk*

To verify whatever deep trenches are sufficient to prevent direct optical crosstalk, we fabricated two different types of 32x1 arrays. Both of them have been manufactured on the wafer W12 and are equipped with deep trenches, which are required for their operation. However, in one of them, each SPAD is surrounded also by the *isolation region*, while in the other it is not. To allow for a fair comparison, the two arrays have the same active area diameter and the same pitch. The space for the isolation region has been obtained in the first type of array by removing the outer guard ring.

To measure the optical crosstalk one array of each type has been mounted (sequentially) in the aforementioned module, which provides a convenient way to operate all the detectors simultaneously in controlled and repeatable conditions. The crosstalk has been evaluated by measuring the correlation between the output pulses generated by a couple of SPADs. In particular, a first SPAD (*emitter*) is connected to the *start* input of a TCSPC instrument (Silena Varro 8k) while the other one (*receiver*) to the *stop* input.

The recorded histogram, once normalized to the number of start-events, represents the distribution of the delays between the firing of the emitter and the *first* successive firing of the receiver, as the Time-to-Amplitude Converter (TAC) can record only the first stop-event after a start-event. By contrast, if every bin of the histogram is normalized to the *effective* number of start-events, i.e. the number of start-events minus the number of events collected in the previous bins, the attained curve represents the distribution of the delays between the firing of the emitter and *any* successive firing of the receiver [64]. In the following we will adopt the latter normalization.

In absence of crosstalk, the firings of the emitter and of the receiver are completely uncorrelated, so we should obtain a uniform (i.e. flat) distribution of the delays. Optical crosstalk causes a deviation from this behavior at nanosecond delays where a photon generated by the emitter SPAD can promptly trigger a correlated avalanche in the receiver. If we subtract the flat contribution due to the uncorrelated events, we obtain the optical crosstalk probability density. Its integral is the optical crosstalk probability.

Fig. 11 represents the optical crosstalk probability as a function of the distance between the emitter and the receiver, measured on two arrays respectively with and without the isolation region. In each curve, the crosstalk dependence on the distance has been attained by fixing the emitter SPAD (roughly at the center of the array) and by changing the receiver SPAD. For each point, the optical crosstalk probability has been calculated following the procedure outlined above.

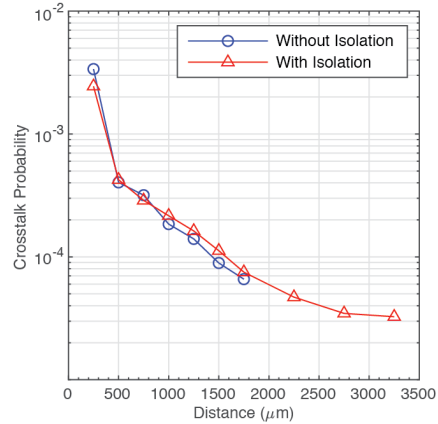

Fig. 11. Optical crosstalk as a function of the distance for RE-SPAD with and without the n+ isolation region (Vov = 20V).

Comparing the two curves of Fig. 11, it is clear that the presence of the isolation region does not have any influence on the value of the optical crosstalk except for the shortest distance (250 µm) where it is slightly reduced. These results suggest that the trench in between two adjacent pixels suppresses most, but not all, of the direct component of optical crosstalk. Conversely, at distances larger than 250 µm, the presence of multiple trenches between the emitter and the receiver is sufficient to completely suppress the direct component of the optical crosstalk; the residual crosstalk we measure is due to the reflection at the bottom of the chip [42] and it is influenced neither by the trenches nor by the isolation. In conclusion, the replacement of the isolation with deep trenches has only a small impact on the optical crosstalk and only for adjacent pixels.

Actually, we cannot even exclude that also at the shorter distance the trench suppresses completely the direct component of the optical crosstalk and the higher value we measure for the optical crosstalk is only due to a random fluctuation. Indeed, we observed a considerable fluctuation in the crosstalk value between adjacent pixels when changing the couple of considered pixels. This is probably due to the roughness of the chip back-surface, which influences the reflection.

## 10. Conclusions

In this paper we demonstrated a silicon technology for the fabrication of SPAD arrays with low timing jitter combined with high detection efficiency at red and near infrared wavelengths. One of the key points for attaining these results is the introduction of deep trenches, which allows the use of thick epitaxial layers while preserving electrical isolation between pixels. We attained a peak detection efficiency exceeding 70% at 650 nm (45% at 800 nm) and a timing jitter of 95 ps FWHM. At the same time other detector's performance, especially the dark count rate and the afterpulsing probability, are not negatively affected by the introduction of deep trenches.

The newly introduced technology enables the fabrication of coarse arrays of SPADs. For example, we demonstrated a 32-pixel linear array with 50 µm active area diameter and 250 µm center-to-center pitch. We think that, despite the small size and small fill factor, this kind of arrays can create exciting opportunities in applications ranging from single molecule spectroscopy to quantum optics.

Currently, the fill factor is limited by the presence of guard rings, while the number of pixels is limited by the need of connecting each detector to an external circuitry through wire-bonding. For the future, we envision a third-generation red-enhanced technology in which these limitations will be overcome with the use of Through the Silicon Vias (TSVs) and with a redesign of the pixel that makes guard rings unnecessary. This would open the way to high density, large arrays of RE-SPAD.

## Funding

National Institute of General Medical Sciences of the National Institutes of Health (NIH) Grant R01 GM095904.
Human Frontier Science Program Grant RGP0061/2019.

## Acknowledgments

The content is solely the responsibility of the authors and does not necessarily represent the official views of the National Institutes of Health.

This work was performed in part at the Cornell NanoScale Science and Technology Facility, a member of the National Nanotechnology Coordinated Infrastructure, which is supported by the National Science Foundation under Grant ECCS-1542081.

## Disclosures

AG, IR: Micro-Photon Devices S.r.l. (C), MG: Micro-Photon Devices S.r.l. (C, I).

## References

1. S. Weiss, "Fluorescence Spectroscopy of Single Biomolecules," Science **283**, 1676–1683 (1999).
2. E. Lerner, T. Cordes, A. Ingargiola, Y. Alhadid, S. Y. Chung, X. Michalet, and S. Weiss, "Toward dynamic structural biology: Two decades of single-molecule Förster resonance energy transfer," Science **359**, eaan1133 (2018).
3. W. Becker, "Fluorescence lifetime imaging - techniques and applications," Journal of Microscopy **247**, 119–136 (2012).
4. A. Pifferi, D. Contini, A. D. Mora, A. Farina, L. Spinelli, and A. Torricelli, "New frontiers in time-domain diffuse optics, a review," J. Biomed. Opt. **21**, 091310 (2016).
5. T. Durduran and A. G. Yodh, "Diffuse correlation spectroscopy for non-invasive, micro-vascular cerebral blood flow measurement," NeuroImage **85**, 51–63 (2014).
6. D. Bronzi, Y. Zou, F. A. Villa, S. Tisa, A. Tosi, and F. Zappa, "Automotive Three-Dimensional Vision Through a Single-Photon Counting SPAD Camera," IEEE Transactions on Intelligent Transportation Systems **17**, 782–795 (2016).
7. H.-K. Lo, M. Curty, and K. Tamaki, "Secure quantum key distribution," Nature Photonics **8**, 595–604 (2014).
8. I. Esmaeil Zadeh, J. W. N. Los, R. B. M. Gourgues, V. Steinmetz, G. Bulgarini, S. M. Dobrovolskiy, V. Zwiller, and S. N. Dorenbos, "Single-photon detectors combining high efficiency, high detection rates, and ultra-high timing resolution," APL Photonics **2**, 111301 (2017).
9. L. Chen, D. Schwarzer, J. A. Lau, V. B. Verma, M. J. Stevens, F. Marsili, R. P. Mirin, S. W. Nam, and A. M. Wodtke, "Ultra-sensitive mid-infrared emission spectrometer with sub-ns temporal resolution," Opt Express **26**, 14859–14868 (2018).
10. E. E. Wollman, V. B. Verma, A. D. Beyer, R. M. Briggs, B. Korzh, J. P. Allmaras, F. Marsili, A. E. Lita, R. P. Mirin, S. W. Nam, and M. D. Shaw, "UV superconducting nanowire single-photon detectors with high efficiency, low noise, and 4 K operating temperature," Opt Express **25**, 26792–26801 (2017).
11. B. Korzh, Q. Y. Zhao, J. P. Allmaras, S. Frasca, T. M. Autry, E. A. Bersin, A. D. Beyer, R. M. Briggs, B. Bumble, M. Colangelo, G. M. Crouch, A. E. Dane, T. Gerrits, A. E. Lita, F. Marsili, G. Moody, C. Peña, E. Ramirez, J. D. Rezac, N. Sinclair, M. J. Stevens, A. E. Velasco, V. B. Verma, E. E. Wollman, S. Xie, D. Zhu, P. D. Hale, M. Spiropulu, K. L. Silverman, R. P. Mirin, S. W. Nam, A. G. Kozorezov, M. D. Shaw, and K. K. Berggren, "Demonstration of sub-3 ps temporal resolution with a superconducting nanowire single-photon detector," Nature Photonics **14**, 250–255 (2020).
12. V. Kotsubo, R. Radebaugh, P. Hendershott, M. Bonczyski, B. Wilson, S. W. Nam, and J. N. Ullom, "Compact 2.2 K Cooling System for Superconducting Nanowire Single Photon Detectors," IEEE Transactions on Applied Superconductivity **27**, 9500405 (2017).

13. N. R. Gemmell, M. Hills, T. Bradshaw, T. Rawlings, B. Green, R. M. Heath, K. Tsimvrakidis, S. Dobrovolskiy, V. Zwiller, S. N. Dorenbos, M. Crook, and R. H. Hadfield, "A miniaturized 4K platform for superconducting infrared photon counting detectors," Superconductor Science and Technology **30**, 11LT01 (2017).
14. S. D. Cova, M. Ghioni, A. L. Lacaita, C. Samori, and F. Zappa, "Avalanche photodiodes and quenching circuits for single-photon detection," Applied Optics **35**, 1956–1976 (1996).
15. F. Ceccarelli, G. Acconcia, A. Gulinatti, M. Ghioni, and I. Rech, "Fully Integrated Active Quenching Circuit Driving Custom-Technology SPADs with 6.2 ns Dead Time," IEEE Photon. Technol. Lett. **31**, 102–105 (2018).
16. F. A. Villa, B. Markovic, S. Bellisai, D. Bronzi, A. Tosi, F. Zappa, S. Tisa, D. Durini, S. Weyers, U. Paschen, and W. Brockherde, "SPAD smart pixel for time-of-flight and time-correlated single-photon counting measurements," IEEE Photonics J. **4**, 795–804 (2012).
17. I. Gyongy, N. Calder, A. Davies, N. A. W. Dutton, R. R. Duncan, C. Rickman, P. Dalgarno, and R. K. Henderson, "A 256 x 256 , 100-kfps, 61% Fill-Factor SPAD Image Sensor for Time-Resolved Microscopy Applications," IEEE Transactions on Electron Devices **65**, 547–554 (2018).
18. A. C. Ulku, C. Bruschini, I. M. Antolovic, Y. Kuo, R. Ankri, S. Weiss, X. Michalet, and E. Charbon, "A 512 × 512 SPAD image sensor with integrated gating for widefield FLIM," IEEE J. Select. Topics Quantum Electron. **25**, 6801212 (2019).
19. X. Michalet, O. H. W. Siegmund, J. V. Vallerga, P. Jelinsky, J. E. Millaud, and S. Weiss, "Detectors for single-molecule fluorescence imaging and spectroscopy," J Mod Optic **54**, 239–281 (2007).
20. S.-K. Liao, W.-Q. Cai, W.-Y. Liu, L. Zhang, Y. Li, J.-G. Ren, J. Yin, Q. Shen, Y. Cao, Z.-P. Li, F.-Z. Li, X.-W. Chen, L.-H. Sun, J.-J. Jia, J.-C. Wu, X.-J. Jiang, J.-F. Wang, Y.-M. Huang, Q. Wang, Y.-L. Zhou, L. Deng, T. Xi, L. Ma, T. Hu, Q. Zhang, Y.-A. Chen, N.-L. Liu, X.-B. Wang, Z.-C. Zhu, C.-Y. Lu, R. Shu, C.-Z. Peng, J.-Y. Wang, and J.-W. Pan, "Satellite-to-ground quantum key distribution," Nature **549**, 43–47 (2017).
21. P. J. Clarke, R. J. Collins, P. A. Hiskett, M.-J. J. García-Martínez, N. J. Krichel, A. McCarthy, M. G. Tanner, J. A. O'Connor, C. M. Natarajan, S. Miki, M. Sasaki, Z. Wang, M. Fujiwara, I. Rech, M. Ghioni, A. Gulinatti, R. H. Hadfield, P. D. Townsend, and G. S. Buller, "Analysis of detector performance in a gigahertz clock rate quantum key distribution system," New J. Phys. **13**, 075008 (2011).
22. E. A. G. Webster, L. A. Grant, and R. K. Henderson, "A High-Performance Single-Photon Avalanche Diode in 130-nm CMOS Imaging Technology," IEEE Electron Device Lett. **33**, 1589–1591 (2012).
23. I. Takai, H. Matsubara, M. Soga, M. Ohta, M. Ogawa, and T. Yamashita, "Single-Photon Avalanche Diode with Enhanced NIR-Sensitivity for Automotive LIDAR Systems," Sensors **16**, 459 (2016).
24. C. Veerappan and E. Charbon, "A Low Dark Count p-i-n Diode Based SPAD in CMOS Technology," IEEE Trans. Electron Devices **63**, 65–71 (2016).
25. M. Sanzaro, P. Gattari, F. A. Villa, A. Tosi, G. Croce, and F. Zappa, "Single-Photon Avalanche Diodes in a 0.16 μm BCD Technology with Sharp Timing Response and Red-Enhanced Sensitivity," IEEE J. Select. Topics Quantum Electron. **24**, 3801209 (2018).
26. Excelitas, "SPCM-AQRH" https://www.excelitas.com/product/spcm-aqrh
27. Laser Components, "COUNT® T" https://www.lasercomponents.com/de-en/product/count-t/
28. H. Dautet, P. Deschamps, B. Dion, A. D. MacGregor, D. MacSween, R. J. McIntyre, C. Trottier, and P. P. Webb, "Photon counting techniques with silicon avalanche photodiodes," Applied Optics **32**, 3894–3900 (1993).
29. B. S. Fong, M. Davies, and P. Deschamps, "Timing resolution and time walk in super low K factor single-photon avalanche diode—measurement and optimization," Journal of Nanophotonics **12**, 1–13 (2018).
30. A. Gulinatti, I. Rech, P. Maccagnani, M. Ghioni, and S. D. Cova, "Large-area avalanche diodes for picosecond time-correlated photon counting," in *Proc. ESSDERC* (2005), pp. 355–358.
31. A. Gulinatti, I. Rech, F. Panzeri, C. Cammi, P. Maccagnani, M. Ghioni, and S. D. Cova, "New silicon SPAD technology for enhanced red-sensitivity, high-resolution timing and system integration," J Mod Optic **59**, 1489–1499 (2012).
32. S. Ates, I. Agha, A. Gulinatti, I. Rech, M. Rakher, A. Badolato, and K. Srinivasan, "Two-Photon Interference Using Background-Free Quantum Frequency Conversion of Single Photons Emitted by an InAs Quantum Dot," Phys. Rev. Lett. **109**, 147405 (2012).
33. M. A. M. Versteegh, M. E. Reimer, K. D. Jöns, D. Dalacu, P. J. Poole, A. Gulinatti, A. Giudice, and V. Zwiller, "Observation of strongly entangled photon pairs from a nanowire quantum dot," Nature Communications **5**, 5298 (2014).
34. D. Bouchet, E. Lhuillier, S. Ithurria, A. Gulinatti, I. Rech, R. Carminati, Y. De Wilde, and V. Krachmalnicoff, "Correlated blinking of fluorescent emitters mediated by single plasmons," Phys. Rev. A **95**, 033828 (2017).
35. A. Ingargiola, E. Lerner, S. Chung, F. Panzeri, A. Gulinatti, I. Rech, M. Ghioni, S. Weiss, and X. Michalet, "Multispot single-molecule FRET: High-throughput analysis of freely diffusing molecules," PLoS ONE **12**, e0175766 (2017).
36. X. Liu, S. Bauer, and A. Velten, "Phasor field diffraction based reconstruction for fast non-line-of-sight imaging systems," Nature Communications **11**, 1645 (2020).
37. G. Blanquer, B. Van Dam, A. Gulinatti, G. Acconcia, Y. De Wilde, I. Izeddin, and V. Krachmalnicoff, "Relocating Single Molecules in Super-Resolved Fluorescence Lifetime Images near a Plasmonic Nanostructure," ACS Photonics **7**, 393–400 (2020).
38. J. Sutin, B. Zimmerman, D. Tyulmankov, D. Tamborini, K. C. Wu, J. Selb, A. Gulinatti, I. Rech, A. Tosi, D. A. Boas, and M. A. Franceschini, "Time-domain diffuse correlation spectroscopy," Optica **3**, 1006–1013 (2016).

39. A. L. Lacaita, M. Ghioni, and S. D. Cova, "Double epitaxy improves single-photon avalanche diode performance," Electron. Lett. **25**, 841–843 (1989).
40. M. Ghioni, A. Gulinatti, I. Rech, F. Zappa, and S. D. Cova, "Progress in Silicon Single-Photon Avalanche Diodes," IEEE J. Select. Topics Quantum Electron. 13, 852–862 (2007).
41. M. Ghioni, A. Gulinatti, I. Rech, P. Maccagnani, and S. D. Cova, "Large-area low-jitter silicon single photon avalanche diodes," in Proc. SPIE **6900**, 69001 (2008).
42. I. Rech, A. Ingargiola, R. Spinelli, I. Labanca, S. Marangoni, M. Ghioni, and S. D. Cova, "Optical crosstalk in single photon avalanche diode arrays: a new complete model," Opt Express **16**, 8381–8394 (2008).
43. A. Gulinatti, F. Panzeri, I. Rech, P. Maccagnani, M. Ghioni, and S. D. Cova, "Planar silicon SPADs with improved photon detection efficiency," in Proc. SPIE **7681**, 76810M (2010).
44. A. Gulinatti, F. Ceccarelli, I. Rech, and M. Ghioni, "Silicon technologies for arrays of Single Photon Avalanche Diodes," in M. A. Itzler and J. C. Campbell, in Proc. SPIE **9858**, 98580A (2016).
45. A. Gulinatti, P. Maccagnani, I. Rech, M. Ghioni, and S. D. Cova, "35 ps time resolution at room temperature with large area single photon avalanche diodes," Electron. Lett. **41**, 272–274 (2005).
46. M. Crotti, I. Rech, A. Gulinatti, and M. Ghioni, "Avalanche current read-out circuit for low-jitter parallel photon timing," Electron. Lett. **49**, 1017–1018 (2013).
47. M. Crotti, I. Rech, G. Acconcia, A. Gulinatti, and M. Ghioni, "A 2-GHz Bandwidth, Integrated Transimpedance Amplifier for Single-Photon Timing Applications," IEEE Trans. VLSI Syst. **23**, 2819–2828 (2015).
48. A. Gallivanoni, I. Rech, and M. Ghioni, "Progress in Quenching Circuits for Single Photon Avalanche Diodes," IEEE Trans. Nucl. Sci. **57**, 3815–3826 (2010).
49. G. Acconcia, M. Ghioni, and I. Rech, "37ps-Precision Time-Resolving Active Quenching Circuit for High-Performance Single Photon Avalanche Diodes," IEEE Photonics J. **10**, 6804713 (2018).
50. M. Assanelli, A. Ingargiola, I. Rech, A. Gulinatti, and M. Ghioni, "Photon-Timing Jitter Dependence on Injection Position in Single-Photon Avalanche Diodes," IEEE J. Quantum Electron. **47**, 151–159 (2011).
51. F. Laemer and A. Schilp, "Method of anisotropically etching silicon," U.S. patent US Patent No. 5,501,893 (1994).
52. S. M. Hu, "Stress from isolation trenches in silicon substrates," JAP **67**, 1092–1101 (1990).
53. W. A. Nevin, K. Somasundram, P. McCann, X. Cao, and S. Byrne, "Factors affecting stress-induced defect generation in trenched SOl for high-voltage applications," in *Semiconductor wafer bonding: science, technology and applications VI* (2002), PV 2001-27, pp 232-241.
54. B. Greenwood, Y. Watanabe, Y. Kanuma, R. Takada, L. Sheng, J. P. Gambino, O. Whear, A. Suhwanov, D. Daniel, S. Menon, D. Price, S. Hose, J. Guo, G. Piatt, and M. Lu, "Gate oxide yield improvement for 0.18μm power semiconductor devices with deep trenches: DP: Discrete and power devices," in Proc. 28th Annual SEMI Advanced Semiconductor Manufacturing Conference (2017), pp. 346–351.
55. A. Ingargiola, M. Segal, A. Gulinatti, I. Rech, I. Labanca, P. Maccagnani, M. Ghioni, S. Weiss, and X. Michalet, "48-spot single-molecule FRET setup with periodic acceptor excitation," The Journal of Chemical Physics **148**, 123304 (2018).
56. A. Gulinatti, I. Rech, P. Maccagnani, and M. Ghioni, "A 48-pixel array of Single Photon Avalanche Diodes for multispot Single Molecule analysis," in Proc. SPIE **8631**, 86311D (2013).
57. G. Vincent, A. Chantre, and D. Bois, "Electric field effect on the thermal emission of traps in semiconductor junctions," JAP **50**, 5484–5487 (1979).
58. F. Ceccarelli, A. Gulinatti, I. Labanca, I. Rech, and M. Ghioni, "Gigacount/Second Photon Detection Module Based on an 8 x 8 Single-Photon Avalanche Diode Array," IEEE Photon. Technol. Lett. **28**, 1002–1005 (2016).
59. A. Giudice, M. Ghioni, R. Biasi, F. Zappa, S. D. Cova, P. Maccagnani, and A. Gulinatti, "High-rate photon counting and picosecond timing with silicon-SPAD based compact detector modules," J Mod Optic **54**, 225–237 (2007).
60. Excelitas, "SPCM-AQRH-TR" https://www.excelitas.com/product/spcm-aqrh-tr
61. R. A. Colyer, G. Scalia, I. Rech, A. Gulinatti, M. Ghioni, S. D. Cova, S. Weiss, and X. Michalet, "High-throughput FCS using an LCOS spatial light modulator and an 8× 1 SPAD array," Biomedical optics express **1**, 1408–1431 (2010).
62. G. Acconcia, I. Rech, I. Labanca, and M. Ghioni, "32ps timing jitter with a fully integrated front end circuit and single photon avalanche diodes," Electron. Lett. **53**, 328–329 (2017).
63. F. Ceccarelli, A. Gulinatti, I. Labanca, M. Ghioni, and I. Rech, "Red-Enhanced Photon Detection Module Featuring a 32x1 Single-Photon Avalanche Diode Array," IEEE Photon. Technol. Lett. **30**, 557–560 (2018).
64. P. B. Coates, "The correction for photon "pile-up" in the measurement of radiative lifetimes," Journal of Physics E: Scientific Instruments **1**, 878–879 (1968).